\documentclass[twocolumn]{aastex631}
\usepackage{balance}

\hypersetup{breaklinks,colorlinks,allcolors=blue}

\begin{document}
\widowpenalty=10000
\clubpenalty=10000
\title{Observations and Radiative Transfer Simulations of the Carbon-rich AGB star V Oph with VLTI/MATISSE\footnote{Based on ESO Program ID 105.208BT, PI Gioia Rau.}}

\author[0009-0007-3944-7298]{Jon Hulberg}
\affiliation{Department of Physics, Catholic University of America, Washington, DC 20064, USA; hulbergjon@gmail.com}
\affiliation{NASA/GSFC Code 667,  Goddard Space Flight Center, Greenbelt, MD 20071, USA}

\author[0000-0002-3042-4539]{Gioia Rau}
\affiliation{National Science Foundation, 2415 Eisenhower Avenue, Alexandria, Virginia 22314, USA} \affiliation{NASA/GSFC Code 667, Goddard Space Flight Center, Greenbelt, MD 20071, USA}

\author[0000-0002-7952-9550]{Markus Wittkowski}
\affiliation{European Southern Observatory, Karl-Schwarzschild-Str. 2, D-85748 Garching bei München, Germany}





\begin{abstract}
Carbon-rich Asymptotic Giant Branch (AGB) stars are among the most important contributors of enriched materials to the interstellar medium due to their strong stellar winds. To fully characterize mass loss on the AGB, it is necessary to determine the distributions of dust and gas around the stars, where the dust begins to condense from the gas, and how this extended atmospheric structure evolves over the pulsational period of the star. We present an analysis of L-band ($2.8-4.2~\mu$m) interferometric observations of the carbon-rich AGB star V Oph made with the MATISSE instrument at the VLTI at the maximum and minimum of the star's visual light curve. Using the radiative transfer software RADMC-3D, we model the circumstellar dust shell, and find stellar radii of 395 and $495~R_{\odot}$ at the two phases, and dust radii of 790 and $742.5~R_{\odot}$ at the two epochs, respectively. By adding C$_2$H$_2$ and HCN gas to the RADMC-3D models, we are able to fit the visibility spectra well, with some deviations at the $3.11~\mu$m feature. Reasons for this deviation and interpretation of the best fitting models are discussed in the text, and we discuss motivations for follow-up imaging observations of V Oph.
\end{abstract}

\section{Introduction} 
\label{sec:intro}

The asymptotic giant branch (AGB) is a short lived but important phase of stellar evolution after the red giant phase experienced by stars with initial masses between 0.8-8 $M_{\odot}$. Although the atmospheres of these stars are initially oxygen rich, thermal pulsing can cause multiple dredge ups of material from the core of the star. After the third dredge up, the carbon to oxygen ratio in the atmosphere can exceed unity \citep{IBEN}. These stars are considered carbon-rich. 

AGB stars undergo physical pulsation, which creates shock waves in their atmosphere that levitate gas. In carbon-rich AGB stars, this creates an extended atmosphere of gas that includes molecules such as CO, HCN, C$_2$H$_2$ \citep{Millar2004}. At low enough density and temperatures, carbon-rich dust can condense out of the gas, creating a circumstellar environment containing species such as silicon carbide (SiC) and amorphous carbon (amC). These dust grains are accelerated by radiation pressure, and through friction drag gas with them, leading to strong stellar winds with mass loss rates up to $10^{-5}~M_{\odot}\cdot yr^{-1}$ (\citealp{Olofsson2004}, \citealp{Hofner2018})  C-rich AGB stars are among the most important contributors of carbon-rich dust and molecules to the interstellar medium. 

Modeling the dusty circumstellar environments and winds of these stars provides valuable insights into the formation of dust and stellar winds. One dimensional hydrodynamic modeling made clear the link between pulsation and dust formation (e.g, \citealp{Bowen1988}, \citealp{Bowen1991}, \citealp{Hoffner2003}, \citealp{DMA}, \citealp{DMA2023}), and current three dimensional star in a box models are being used to understand the clumpy, asymmetric formation of dust around AGB stars as described in \citet{FH2023} and \citet{ahmad2023}.

 For the task of modeling interferometric observations of individual stars, publicly available radiative transfer codes such as RADMC-3D \cite{Dullemond2012} are ideal. They allow the user to specify an environment of dust and gas around a source star that can be specifically tailored to a desired target. Comparing these models to high angular resolution observations from optical interferometry provides valuable insights into their circumstellar environment.

 V Ophiuchi (hereon refered to as V Oph) is a carbon-rich AGB star located at a distance of $741~\pm~14$~pc \citep{GaiaDR3}. Optical and mid-infrared (IR) interferometric observations provide a high-angular resolution view of its extended atmosphere and circumstellar environment, making it a good target to use in studies of AGB stellar winds. Previous observations in the N-band with the Very Large Telescope Interferometer's (VLTI) MID-infrared Interferometric instrument (MIDI) \citep{MIDI} were used with radiative transfer modeling to determine the star was surrounded by two thin layers of C2H2 and HCN, as well as a dusty shell \citep{OH7}. Further modeling by \cite{RAU2019} compared the MIDI observations to the grid of C-rich DARWIN model atmospheres from \cite{DMA}, concluding the models reproduced photometric data well, but did not fully reproduce the high-angular resolution MIDI observations. 

 We report IR interferometric observations of V Oph with the new MultiAperTure mid-Infrared SpectroScopic Experiment (MATISSE) (\citealp{MATISSE})at the VLTI (ESO program ID
105.20BT, PI Rau). MATISSE performs measurements in the L-,M-, and N-bands, making it an ideal instrument to probe the locations of carbon rich molecules such as HCN and C$_2$H$_2$ (L- and M-bands), and dust grains (N-band). We use the radiative transfer software package RADMC-3D to model the distribution of carbon-rich dust and gas in the extended atmosphere. We created a grid of models in order to study the time dependence of the atmospheric structure. The goals of this work are to constrain the distribution of gas and dust, to understand where dust begins to form, and how the stellar structure varies with pulsational phase.

 The article is structured as follows: The observations and data reduction are described in Section \ref{sec:Observations}, the radiative transfer modeling and fitting is described in Section \ref{sec:Modeling}, and the results are discussed in Sections \ref{sec:Results} and \ref{sec:Discussion}.
 
 \section{Observations and Data Reduction}\label{sec:Observations}

\subsection{Photometry}
\begin{figure}
\plotone{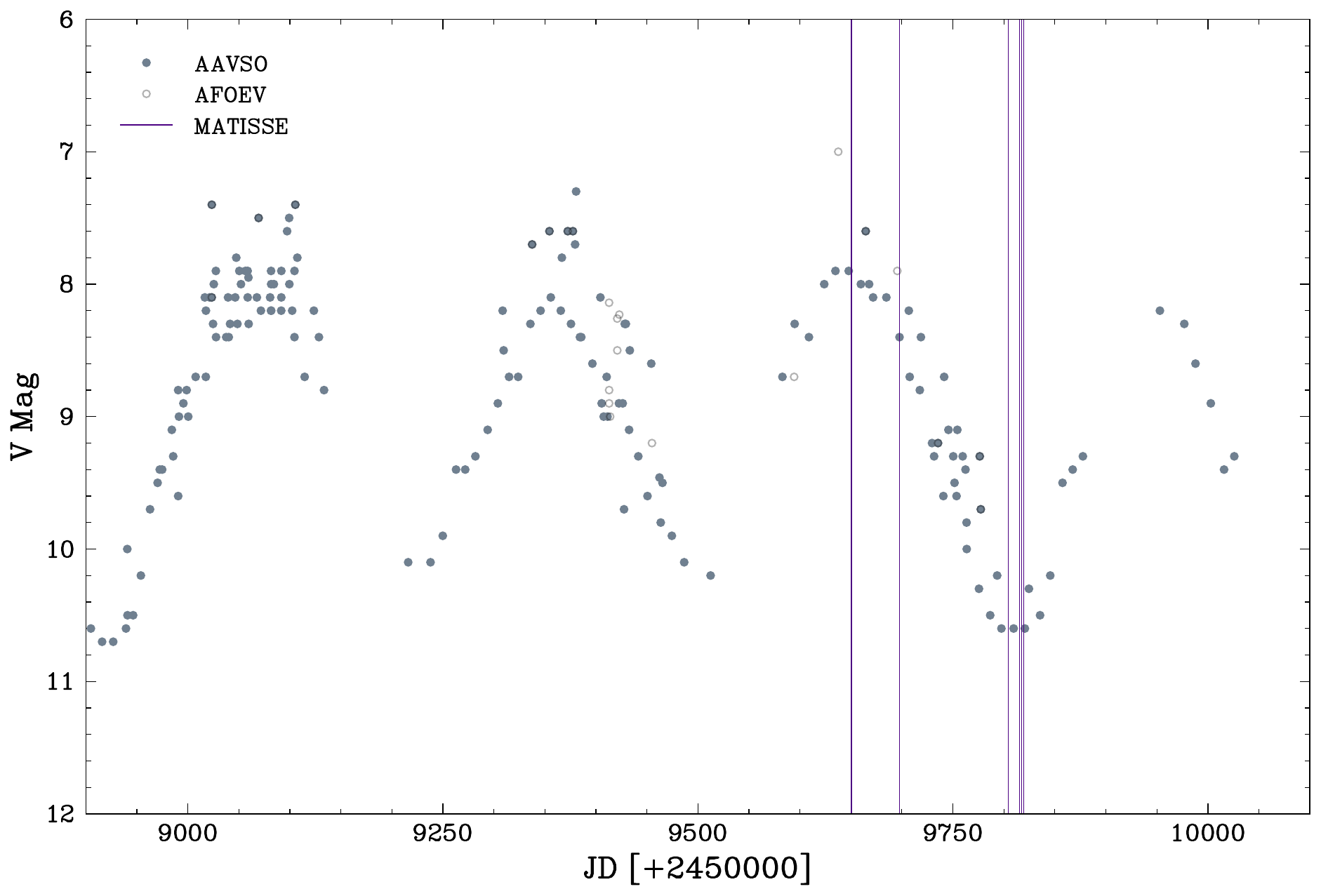}
    \caption{Visual light curve of AAVSO and AFOEV data. The time of MATISSE observations are indicated by vertical purple lines.}
    \label{fig:lightcurve}
\end{figure}

Figure \ref{fig:lightcurve} shows a light curve constructed using publicly available V-band photometric data from the American Association of Variable Star Observers (AAVSO - \citealp{AAVSO}) and the French Association of Variable Star Observers (AFOEV - \citealp{AFOEV}). We calculated the phase listed in Table~\ref{tab:obs} for each MATISSE observation by fitting a sine curve to the AAVSO data to find the period of $300.4~\pm~0.8$ days and time of maximum V-band light closest to our MATISSE observations at at $2,459,661~\pm~2$ JD.  

\subsection{Interferometetry}
MATISSE combines light in the L ($2.8-4.2~\mu$m), M ($4.5-4.0~\mu$m), and N ($8-13~\mu$m) infrared bands from the four VLTI unit or auxiliary telescopes (ATs), providing 6 visibility measurements and three closure phase measurements per observation. 
The interferometric observations (ESO program ID 105.20BT, PI Rau) were observed with the MATISSE instrument in the L- and N-Band with low spectral resolution (R=34/30) on the nights of 2022-03-12, 2022-04-28, and 2022-08-26, and 2022-08-28.  Additional observations were made on 2022-08-13 in the M-band with medium spectral resolution (R=506), and on 2022-08-24 in the L-band with medium resolution (R=506). Table \ref{tab:obs} shows the details of the observations; They are marked as vertical purple lines in Figure \ref{fig:lightcurve}. The observations cover both maximum and minimum light on the visual light curve.
Each observation was made using the standard CAL-SCI-CAL block. The L-band observations were taken in the SCIPHOT mode, i.e, the source photometry and fringes are measured simultaneously. Due to the large thermal background in the N-band, it is necessary to observe photometry and fringes sequentially (HIGHSENS mode). The combination of these two settings for the two bands is called HYBRID mode and was the only option offered for MATISSE at the time of the observations (\citealp{manual}). The lone M-band observation was also made in HIGHSENS mode.

\begin{deluxetable*}{lllllllc}[t]\label{tab:obs}
\tablehead{\colhead{Observation}&\colhead{Julian Date}&\colhead{}&\colhead{Binned}&\colhead{Array}&\colhead{Resolving Power }&\colhead{}&\colhead{Science Seeing} \\[-0.2cm]
\colhead{date}&\colhead{(+2450000)}&\colhead{Phase}&\colhead{phase}&\colhead{configuration}&\colhead{[$\frac{\lambda}{\Delta \lambda}$]}&\colhead{Calibrators}&\colhead{[$^{\prime\prime}$]}}
\tablecaption{MATISSE observations of V Oph.}
\startdata
        2022-03-12 &9650.5& 0.96 & 0.04& Small &34(L)/30(N)& HD 147647 & 0.81\\
        2022-04-28 &9697.5& 0.12& 0.04& Small &34(L)/30(N)& HD 146051 & 0.56 \\
        2022-08-13 &9804.5& 0.47& 0.51 &Small &506 (M)&  HD 163917/HD 175775 & 0.89\\
        2022-08-24 &9815.5& 0.51 & 0.51&Small &506 (L) & HD 148633/HD 147647 & 0.54 \\
        2022-08-26 &9817.5& 0.52& 0.51& Medium &34(L)/30(N)& HD 147647 & 0.71 \\
        2022-08-28 &9819.5& 0.53& 0.51& Large &34(L)/30(N)& HD 146051 & 0.40 \\
\enddata
\tablecomments{ESO Program ID 105.208BT, PI Gioia Rau.}
\end{deluxetable*}

\subsection{Data Reduction}\label{sec:reduction}
We processed using version 2.0.0 of the MATISSE data reduction pipeline\footnote{\url{https://www.eso.org/sci/software/pipelines/matisse/matisse-pipe-recipes.html}} (\citealp{Millour}) Reflex graphical interface (\citealp{reflex}). The pipeline calculates incoherent estimators, squared visibilities and closure phases from the Fourier transformed interferogram. We reduced the N-band data using a spectral binning of 11 and coherence time of 0.3 ms and reduced the L-band data using the standard spectral binning of 5. 

\begin{figure*}[htb!]
    \plotone{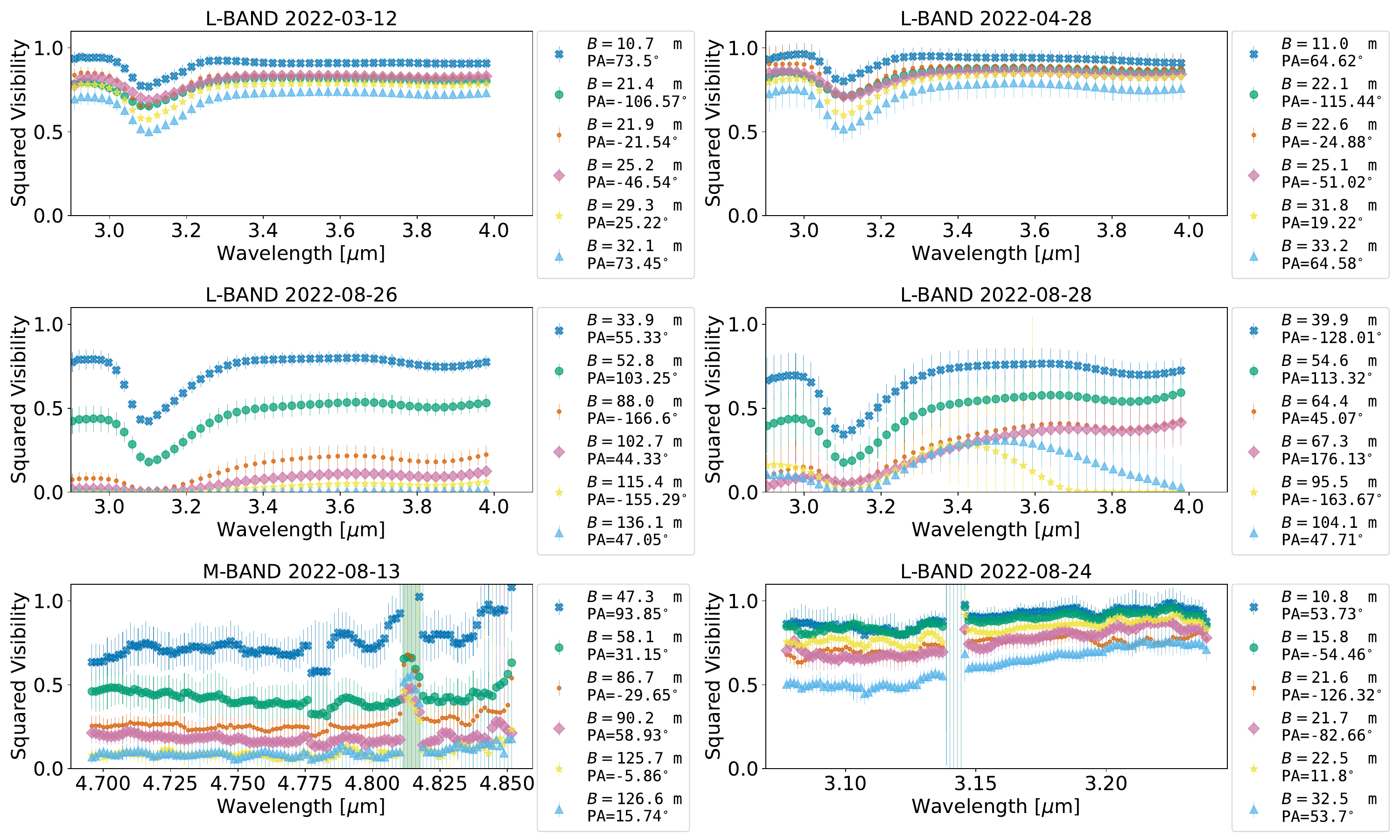}
    \caption{L-band and M-band calibrated observations, the different colors show data for  different projected baselines and position angles. A known telluric line in the M-band is shaded in green.}
    \label{fig:Ldata}
\end{figure*}

\begin{figure*}[htb!]
    \plotone{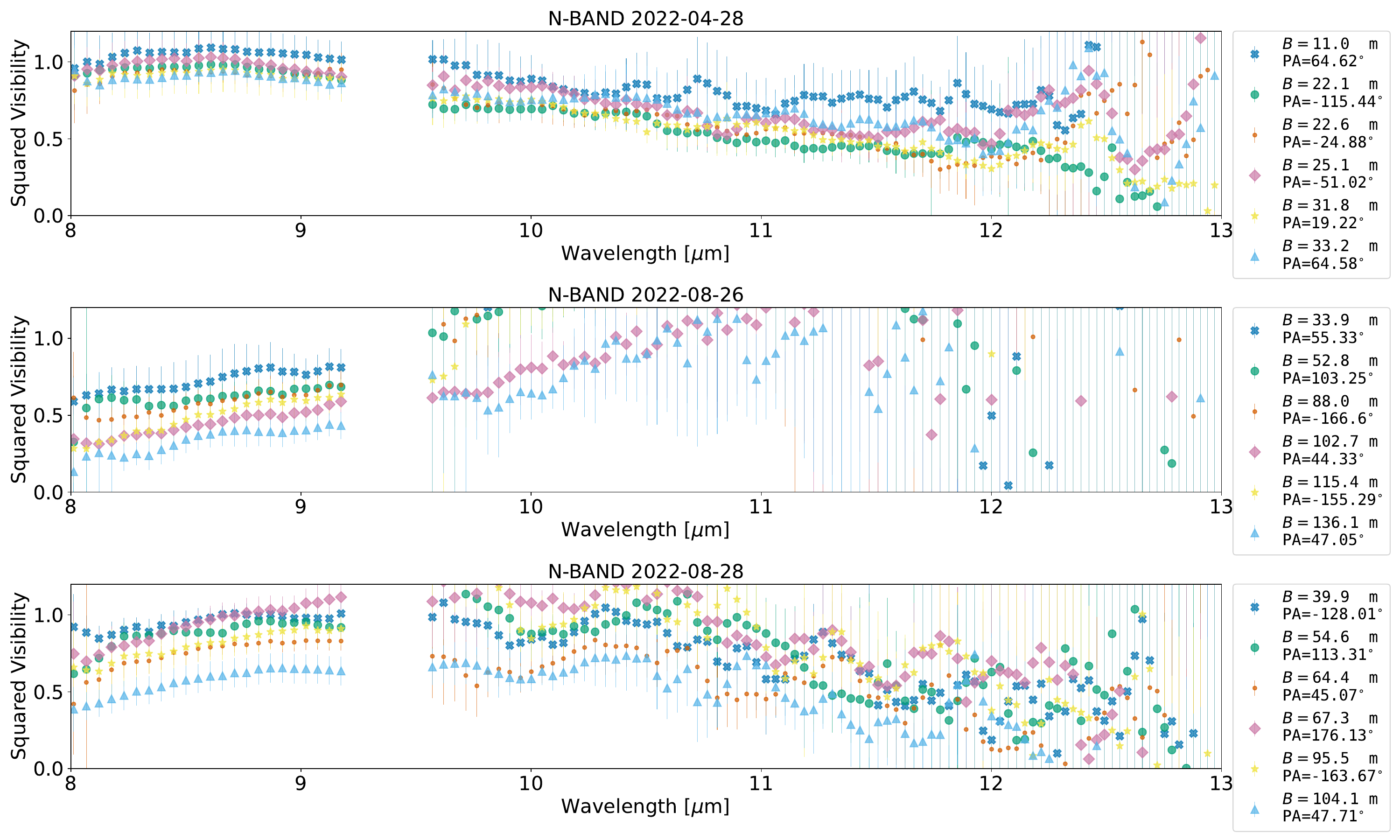}
    \caption{Same as Figure~\ref{fig:Ldata}, for the N-band. We do not include data from the night of 2022-03-12 because it suffered from calibration issues causing it to appear over-resolved. The gap around $9.5~\mu$m is explained by the MATISSE data reduction pipeline adding flags around a telluric line.}
    \label{fig:Ndata}
\end{figure*}

We calibrated the visibilities for atmospheric and instrumental visibility losses using the interferometric transfer function. The transfer function is the ratio between the measured calibrator visibility and the known calibrator visibility based on an adopted angular size of the calibrator (Equation \ref{eq:TF}). The transfer function was divided into the science visibilities to calibrate them. The L-band and N-band calibrated visibilities are shown in Figures~\ref{fig:Ldata} and \ref{fig:Ndata}.

\begin{equation}
    TF^2 = \frac{V^2_{calibrator}}{V^2_{known}}
    \label{eq:TF}
\end{equation}

\subsection{Visibility Data \& Binning}\label{subsec:visdata}

Figure~\ref{fig:Ldata} shows the L-band observations. The visibility profiles are characterized by a mostly flat visibility profile, with the exception of a broad drop in visibility around 3.1 $\mu m$ that can be attributed to emission from HCN and C$_2$H$_2$ gas. These gas species also cause smaller drop in visibility that can be seen around 3.8 $\mu m$ at longer projected baselines on the nights of 2022-08-26 and 2022-08-28. On the night of 2022-08-28, the longest projected baselines of 95.5 m and 104.1 m show a sharp drop in visibility beginning around 3.5 $\mu m$. Unfortunately, the star used as calibrator for that night shows deviation from a uniform disk visibility for those baselines, which makes the transfer function estimate for those baselines poor. As such we exclude these two baselines from the analysis presented in Section \ref{sec:Modeling}. Figure~\ref{fig:Ldata} also shows the lone M-band observation on the night of 2022-08-13. 

The N-band data shown in Figure~\ref{fig:Ndata} is characterized by a flat or increasing visibility profile from $8-9~\mu$m, followed by a gap in the data where the reduction workflow flags wavelengths that are affected by a telluric line in the atmosphere. At longer wavelengths, the data suffers from severe noise introduced in the calibration process. Due to the noise, it is difficult to detect the effects of SiC emission or gas absorption in the visibilities at long wavelengths. The long wavelength noise is expected given the MATISSE flux limits. The MATISSE instrument page\footnote{\url{https://www.eso.org/sci/facilities/paranal/instruments/matisse/inst.html}} provides flux limits required to achieve a precision of 0.1 in visibility for three sub-bands in the N-band.  These bands are defined as the N1-band ($8.5~\pm~0.4~\mu$m), N2-band ($10.5~\pm~0.5~\mu$m), and N3-band ($11.5~\pm~0.5~\mu$m). The flux limits for the auxiliary telescopes are $17~Jy$, $32~Jy$, and $45~Jy$. The $12~\mu$m flux of V Oph is $29.0~Jy$ in the IRAS Catalogue of Point Sources \citep{iras}, and $25.4~Jy$ in the ALLWISE Source Catalogue \citep{WISE}. Although the flux of V Oph is above the N1-band flux limit, it lies below the N2- and N3-band flux limits. This explains the general trend of decreasing precision with increasing wavelength in the N-band. The N-band visibility levels from the night of 2022-03-12 were much lower than the visibility levels at similar baselines on the night of 2022-04-28, suggesting an unrealistic uniform disk diameter of $45\pm1$~mas. We note the calibrator (HD 147647) has the infrared (IR) excess, IR extent, and mid-infrared (MIR) variability flags in the Mid-infrared stellar Diameters and Fluxes compilation Catalogue (\citealp{Cruzalebes2019}). This indicates it may be a poor calibrator in the N-band, so we elected to discard the N-band observations from the night of 2022-03-12.

\begin{figure}
    \plotone{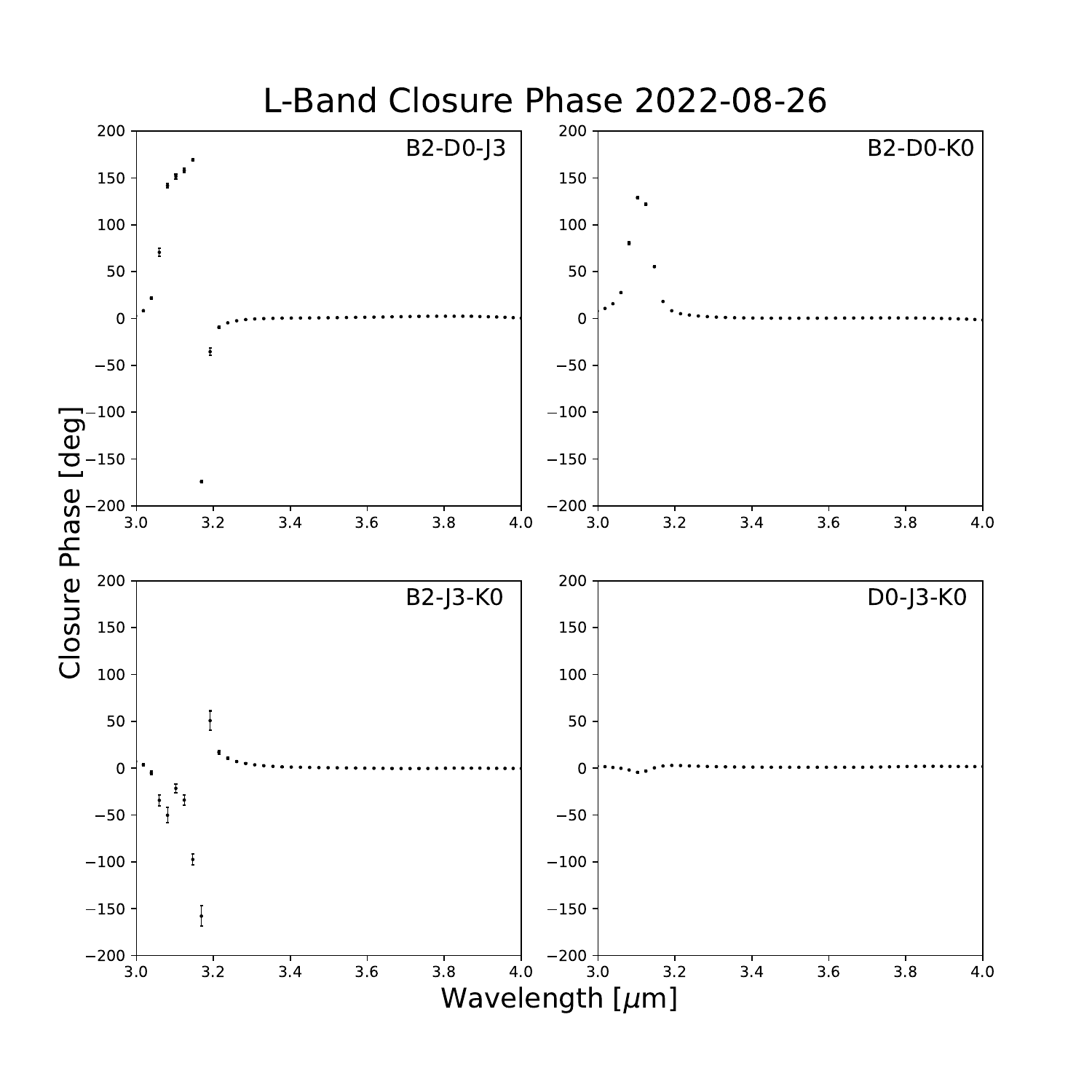}
    \caption{L-band closure phases from the night of 2022-08-26 labeled by baseline triangle.}
    \label{fig:closurephase}
\end{figure}

The two observations on 2022-03-12 ($\phi=0.96~\pm~0.007$) and 2022-04-28 ($\phi=0.12~\pm~0.007$)  were taken close to maximum visual light in the light curve shown in Figure~\ref{fig:lightcurve}. Although their difference in phase is 0.16, direct comparison of their visibility profiles for corresponding baselines and spectral channels shows a median difference of $1.3~\sigma$, where $\sigma$ is difference divided by the uncertainty in the visibility measurement on 2022-03-12. The maximum deviation is only $3.0~\sigma$. We conclude the observations from these nights are consistent with each other and elected to bin the observations together with an average phase of $\phi=0.04~\pm~0.005$. The remaining four observations were taken between 2022-08-13 and 2022-08-28. The 15 day difference between the first and last of these four observations corresponds to a difference in phase of 0.05. Because of this small difference, we the grouped these four observations into a second bin with an average phase of $\phi=0.51~\pm~0.004$. We performed most of the modeling described in the rest of this paper on the data grouped into these two epochs of 0.04 and 0.51, referred to as the maximum and minimum visual light curve epochs. 
The closure phase in the L-band are generally flat with values near zero or $\pm~180^{\circ}$ with some deviation from centro-symmetry at the 3.11 $\mu m$ feature. Figure \ref{fig:closurephase} shows the closure phases from the night with the strongest evidence of such deviations, 2022-08-26. 

\section{Modeling}\label{sec:Modeling}
\subsection{Geometrical Modeling}\label{sec:geom}
\begin{figure*}
    \plotone{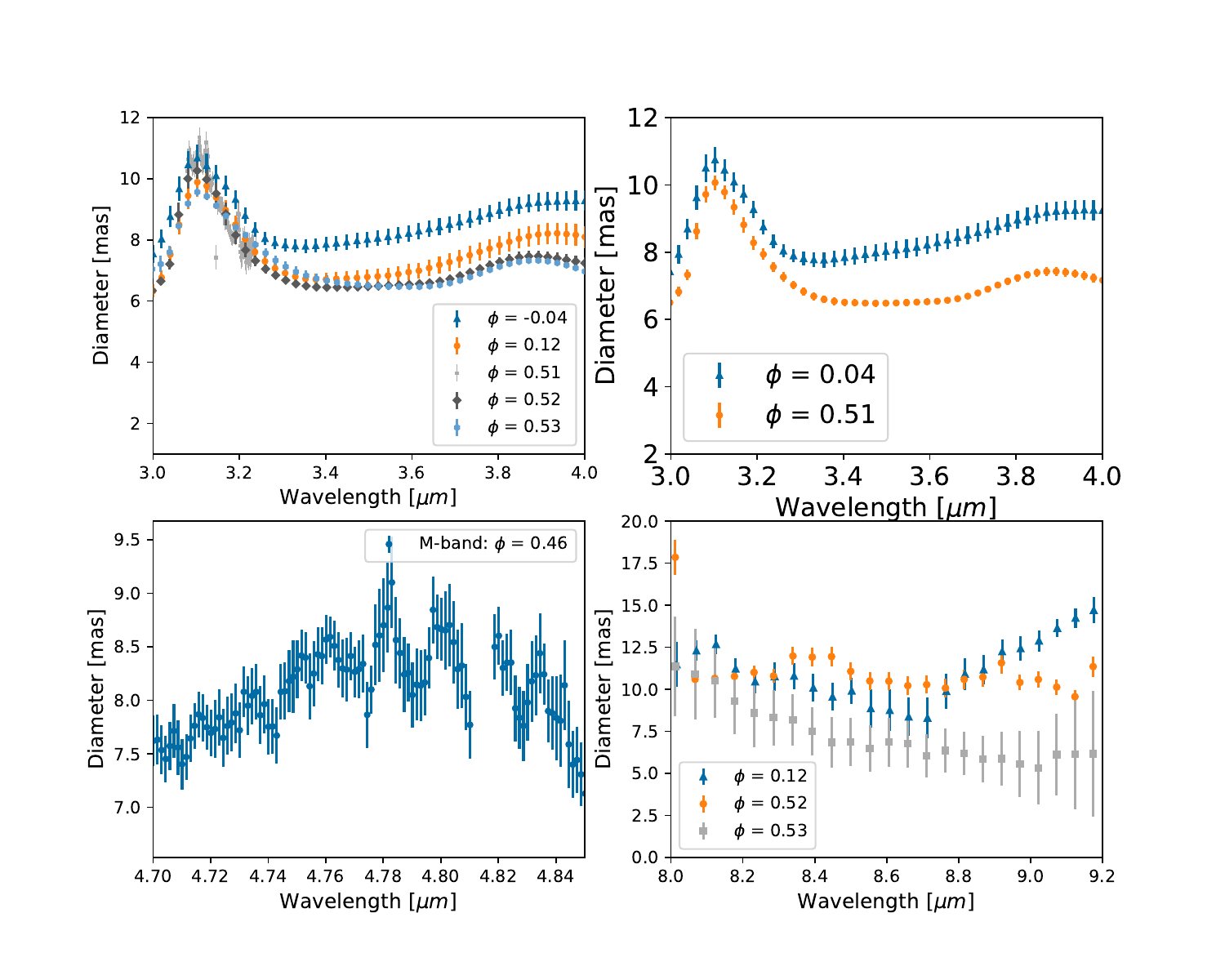}
    \caption{PMOIRED uniform disk diameters vs. MATISSE spectral channel. Top left - Uniform disk diameters vs. spectral channel at phases of -0.04, 0.12, 0.51, 0.52, and 0.53 represented by dark blue, orange, gray, black, and light blue points respectively. Top right - L-band uniform disk diameters for binned phases of 0.04 and 0.51. Bottom left - Uniform disk diameters for the sole M-band observation. The missing data points around 4.81~$\mu m$ correspond to a telluric feature. Bottom right - N-band uniform disk diameters vs. spectral channel for the phases of 0.12, 0.52, and 0.53 represented by dark blue, orange, gray, and black points respectively. The N-band fits are restricted to wavelengths between $8.0-9.2~\mu$m.}
    \label{fig:pmoirednobin}
\end{figure*}

To estimate the angular size in the L-band and N-band, we used the Python package PMOIRED (\citealp{PMOIRED}) to fit the visibility data to a uniform disk model. Figure~\ref{fig:pmoirednobin} shows derived angular diameters per MATISSE spectral channel in the L-band, as well as in the N-band prior to the telluric feature around $9.2~\mu$m. We do not show fits at longer wavelengths in the N-band because many of the fits had trouble converging due to the noise level. The L-band visibilities deviate from a uniform disk at projected baselines larger than $90~m$, so we exclude these baseline from the fit. We also show uniform disk fits to the low resolution L-band observations grouped into their bins of $\phi=0.04$ and $\phi=0.51$ with the largest projected baselines removed. A fit to the sole M-band observation is also shown. Table \ref{tab:UDtable} shows the uniform disk diameters at wavelengths of $3.1$, $3.4$, and $8.5~\mu$m.

\begin{deluxetable*}{rrlllll}\label{tab:UDtable}
\tablehead{\colhead{}&\colhead{$\Theta_{3.1}$}&\colhead{$\Theta_{3.4}$}&\colhead{$\Theta_{8.5}$}&\colhead{}&\colhead{}&\colhead{}\\[-0.2cm]
\colhead{Phase}&\colhead{[mas]}&\colhead{[mas]}&\colhead{[mas]}&\colhead{$\chi^2_{3.1}$}&\colhead{$\chi^2_{3.4}$}&\colhead{$\chi^2_{8.5}$}}
\tablecaption{Uniform disk diameters and reduced $\chi^2$ at $3.1$, $3.4$, and $8.5~\mu$m for the binned phases of 0.04 and 0.51.}
\startdata
        0.04 & $10.7~\pm~0.4$& $7.8~\pm~0.2$ &$9.9~\pm~0.88$& 3.3 & 0.8& 0.1\\
        0.51 & $10.1~\pm~0.2$& $6.5~\pm~0.1$ &$8.5~\pm~0.9$& 1.0 & 0.1 & 1.6\\      
\enddata
\end{deluxetable*}

\subsection{Radiative Transfer Modeling}
 To model both the dust shell and molecular layers around the star, we used the radiative transfer software RADMC-3D. This software allows the user to specify an arbitrary distribution of gas and dust around a central source. The dust temperature is calculated using a random walk Monte-Carlo procedure as described in \cite{BjorkmanWood2001} and \cite{Lucy1999} to simulate dust scattering and absorption. RADMC-3D calculates images and fluxes using ray tracing. When molecular species are included in addition to dust, the line radiative transfer is computed based on user provided molecular line data with level populations calculated assuming local thermal equilibrium (LTE). Because of the density and width of C$_2$H$_2$ and HCN absorption lines in the near infrared, a spectral resolution on the order of $0.1~\mathring{A}$ is needed to accurately simulate the contribution of molecular layers around the star. To limit computation time, we implemented a two step modeling process where first the central source and dust shell modeled at a low spectral resolution without the contribution of the molecular layers, allowing us to create a grid of models with different stellar and dust shell parameters. After finding the best fitting dust shell only model at both epochs, we performed high spectral resolution simulations ($0.15~\mathring{A}$) of the molecular layers. The dust shell only modeling is described in Section~\ref{subsec:dustshell} and the gas shell modeling is described in Section~\ref{subsec:gasshell}.
\subsection{Dust Shell and Stellar Parameters}\label{subsec:dustshell}
In the first step of the modeling, we produced a dust shell only grid made of 1750 different models. Each model is fully described by a set of 8 parameters related to the star, and dust shell. The stellar parameters are the effective temperature $T_{eff}$ of the blackbody source and the stellar radius $R_*$. The dust shell parameters are the dust condensation radius $R_d$, the amC mass density in each cell of the simulation $\rho_{amC}$, the SiC mass density in each cell $\rho_{SiC}$, the dust grain size distribution for both dust species $n(a)_{amC/SiC}$, and the outer radius of the simulation $R_{out}$. We fixed the effective temperature of the source at 2600 K for all models based on the results of \cite{RAU2019}, and parameterized the dust densities with a spherically symmetric $r^{-2}$ density profile composed of 90\% amC and 10\% SiC dust (composition following \citealp{Rau2015}, \citealp{Rau2017}). With this parametrization, the dust mass is determined by only one parameter: the total dust density at the dust radius.  We fixed the grain size distribution to $0.2~\mu$m following the modeling of \cite{RAU2019} and calculated the dust opacities using the software package optool (\citealp{optool}) with optical constants from \cite{Zubko}\footnote{URL to the amC optical constants in the optool github: \url{https://github.com/cdominik/optool/blob/master/lnk_data/c-z-Zubko1996.lnk}} and \cite{Pegourie}\footnote{URL to the SiC optical constants that we provided to optool: \url{https://github.com/ivezic/dusty/blob/master/data/stnd_dust_lib/SiC-peg.nk}}.
We chose a grid of photospheric radius values from 330 $R_\odot$ to 500 $R_\odot$ with a spacing of 5 $R_\odot$ and a grid of dust radius values between 1.5 and 4.0 $R_*$ with a spacing of 0.5 $R_*$. We selected a broad range of photospheric radii due to the lack of contemporaneous spectral energy distribution data, which could have been used to constrain the stellar radius. We set the dust radius spacing  to a relatively low resolution of 0.5 $R_*$ because the N-band observations could not be utilized to constrain the dust shell, as discussed in Section \ref{sec:Observations}, due to the quality issues with the N-band data. We note that the stellar radius resolution of 0.5 $R_*$ is the same used in \citet{OH7} for the grid search modeling of V Oph MIDI observations.  These parameter ranges encompass the result of the modeling of both \cite{OH7} and \cite{RAU2019}. The dust mass was varied by changing the dust density at the a radius of $R_d$ with total dust masses ranging between $1 \times 10^{-10}$ to $1\times 10^{-8}$ $M_\odot$. We chose these dust densities and their resulting dust masses based on initial modeling where we matched the model visibility profiles and a spectral energy distribution to the MATISSE data as well as the N-band spectra from \citet{OH7} by adjusting the dust mass by hand. Because of the approximate nature of this initial modeling, we conservatively chose a wide range of dust masses around the initial modeling values. We fixed the outer radius of each model at $140$~AU.

\begin{deluxetable*}{llllll}[bth!]\label{tab:dustshellparameters}
\tablehead{\colhead{}&\colhead{Photospheric Radius}&\colhead{Dust Radius}&\colhead{Dust Mass}&\colhead{L-Band}&\colhead{N-Band}\\[-0.2cm]
\colhead{Phase}&\colhead{($R_{\odot}$)}&\colhead{($R_*$)}&\colhead{($M_{\odot}$)}&\colhead{$\chi^2_{red}$}&\colhead{$\chi^2_{red}$}}
\tablecaption{Best fitting star + dust shell models.}
\startdata
        0.04 & 395 &2.0& $1.4 \times 10^{-9}$ & 0.40 & 39.4\\
        0.51 & 495 &1.5& $6.5 \times 10^{-10}$ &0.18 & 2.1\\
\enddata
\end{deluxetable*}

The output of RADMC-3D for each model is a radial intensity profile in the L-band. From these profiles, we calculated synthetic visibility profiles using a Hankel transform (\cite{Tango2000}, \cite{TD2002}), which is the simplification of a Fourier transformation in the case of radial symmetry. The visibility calculations were performed using a distance of $641\pm14$~pc from the Gaia Data Release 3 (\citealp{GaiaDR3}).  The $\chi^2$ of each model was evaluated at the binned phases of $\phi=0.04$ and $0.51$ over the wavelength range of 3.4-3.6 $\mu$m where the contribution of C$_2$H$_2$ and HCN is minimal. The best fitting models are summarized in Table \ref{tab:dustshellparameters}. The N-band visibilities were not used to find the best fitting model because the severe noise levels discussed in Section \ref{subsec:visdata} would lead to inaccurate results. To compare the models to the N-band data, we calculated N-band visibility profiles for the best fitting model at both binned phases. The N-band $\chi^2$ is reported together with the L-band $\chi^2$ in Table \ref{tab:dustshellparameters}. 
\subsection{Molecular layers}\label{subsec:gasshell}
In the second step of the modeling process, we added a molecular shell to the best-fitting models obtained in the first step. In the circumstellar environments of AGB stars, both gas and dust can coexist at various radii.   \citet{OH7} modeled a distinct molecular shell located close to the photosphere of AGB stars; we adopt a similar approach, modeling a pronounced molecular gas shell near the photosphere, surrounded by a dust-only CSE. In addition to the dust shell parameters described in Section~\ref{subsec:dustshell}, the dust+gas models have 6 additional parameters that describe the gas shell. The number density distributions of both C$_2$H$_2$ and HCN, the outer radius of the most pronounced gas shell, the temperature of both species of gas, and the microturbulent velocity of the gas $v_{kms}$. We added the molecular shell of C$_2$H$_2$ and HCN to the best fitting dust shell only model at each phase using molecular line data compiled from the HITRAN \citep{HITRAN} database using the Python module HAPI \citep{HAPI}. A grid search to optimize the parameters of the gas shell would be prohibitively expensive in terms of computation time. Therefore, we fine tuned the parameters of the gas shell with simulations performed at a lower spectral resolution. We started by performing simulations with a resolution of approximately $2~\mathring{A}$. We found that simulations performed at this resolution match closely the high resolution results, so a lower spectral resolution can be used to reduce computation time if the final result is verified at the full spectral resolution. The initial values of the C$_2$H$_2$ and HCN densities were based on the gas density at $2.5~R_*$ given \citet{Cherchneff2012} as well as LTE fractional densities for a carbon star given in \citet{Millar2004}. A $r^{-2}$ density profiles was assumed up to the outer radius of the gas shell.   The gas radius values were chosen based on the modeling of \cite{RAU2019}. We adopted a fixed gas temperature of $1200~K$ for both species of gas because it fell within the range of gas temperatures found by \cite{OH7} and produced results that agree well with the measured visibilities. We held the microturbulent velocity fixed at $v_{kms}=2.5~km/s$ following \citet{DMA}.

To fine tune the parameters, we adjusted the gas densities and gas shell radius for the  $\phi=0.04$ and $\phi=0.51$ models until a $\chi^2$ close to one was achieved. The $\chi^2$ was calculated using the visibility data between $3.0-4.0~\mu$m, and a conservative fractional error of 10\% was adopted following \cite{Rau2015}. Additionally, at large projected baselines, the visibilities are almost resolved out and have very small uncertainties. We imposed a flat minimum error of 0.01 in addition to the 10\% fractional minimum so the fitting process did not overweight these baselines compared to the rest of the data. When calculating the $\chi^2$, we did not include the baselines of $95.5$ and $104~m$ from the night of 2022-08-28 as explained in in Section \ref{sec:Observations}.

\begin{figure*}
    \plotone{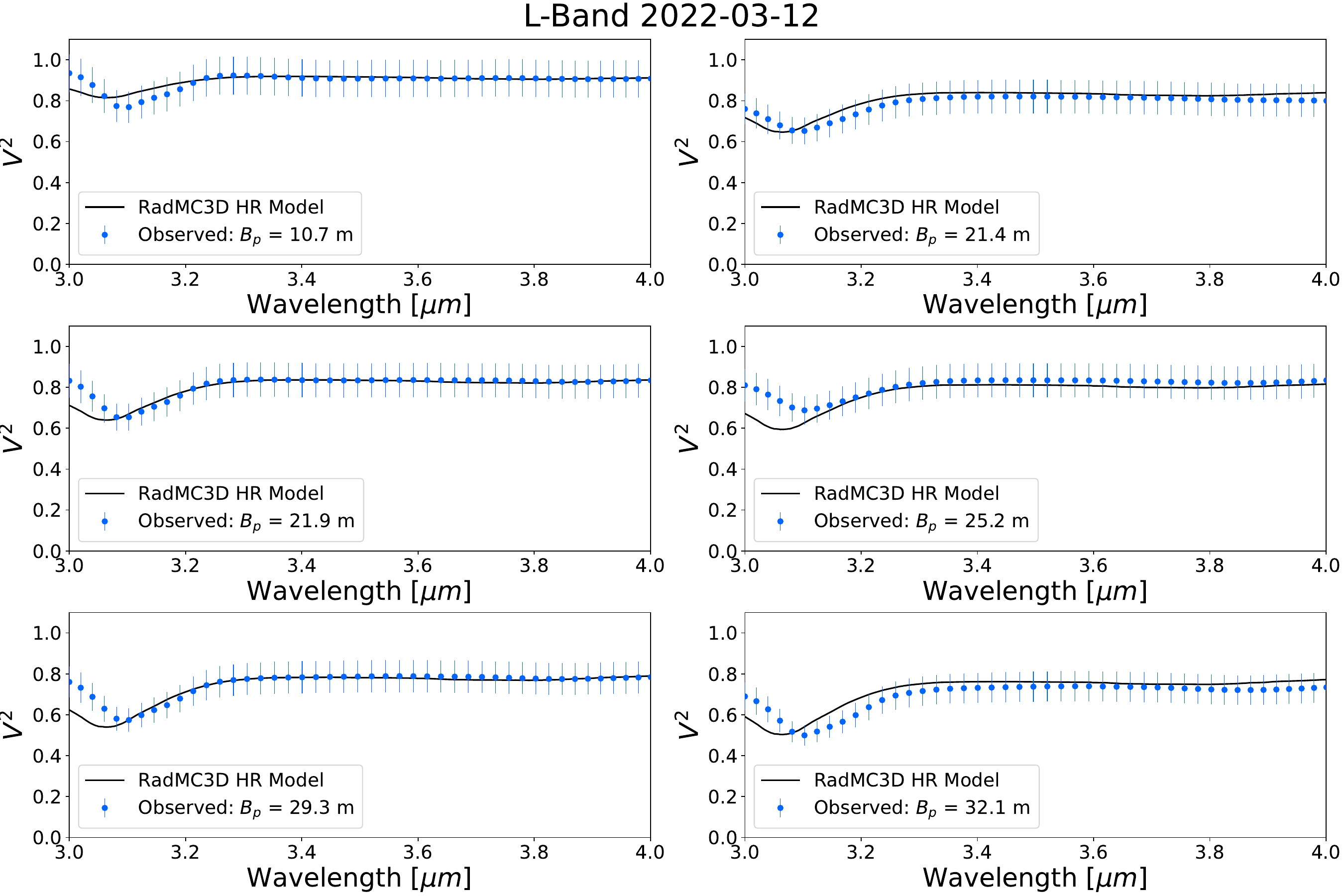}
    \caption{Synthetic RADMC3D-calculated (black lines) and observed (blue dots) L-band visibility profiles for the maximum light epoch on the night of of 2022-03-12.}
    \label{fig:gasvismax}
\end{figure*}

\begin{figure*}
    \plotone{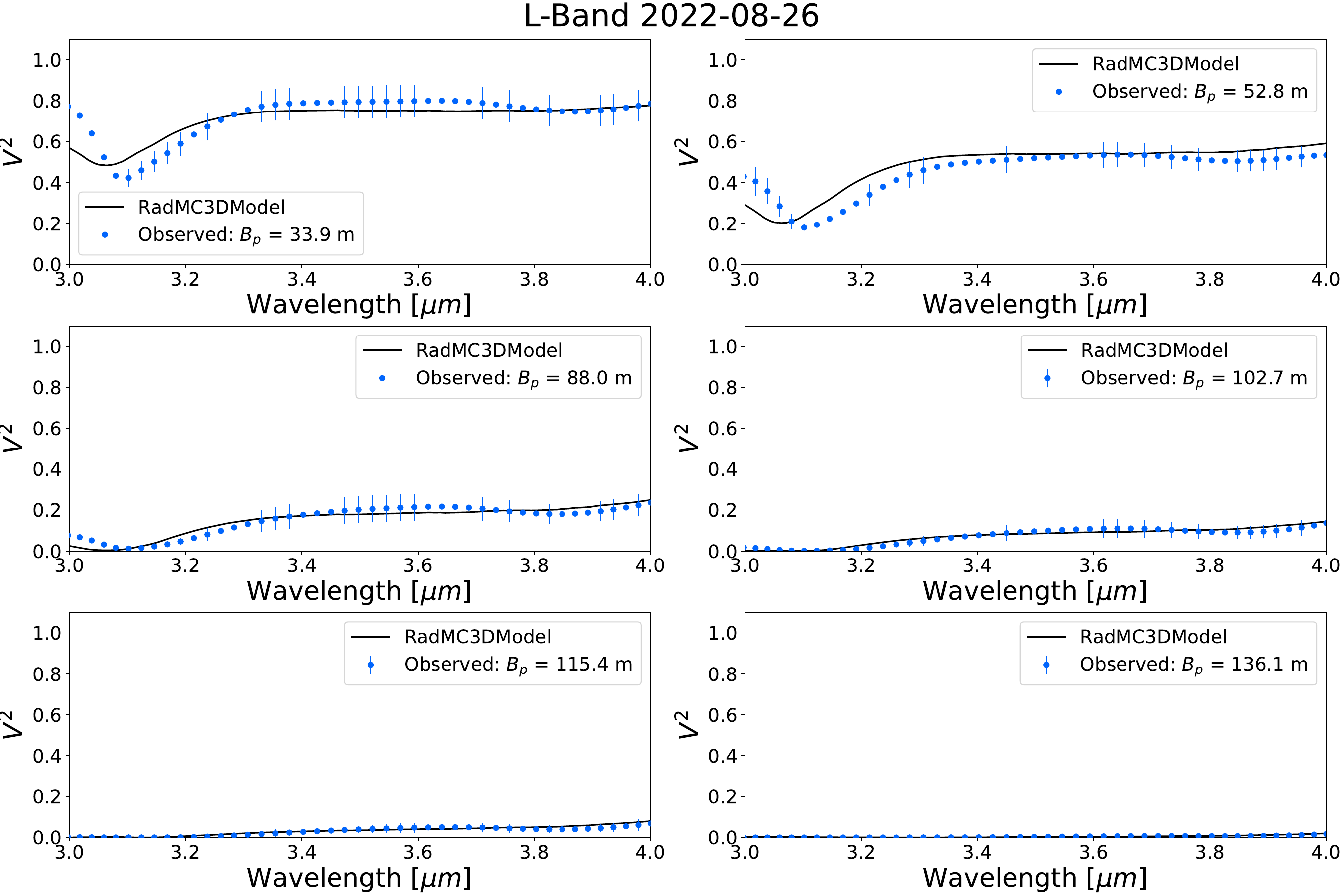}
    \caption{Same as Figure~\ref{fig:gasvismax} but for the night of 2022-08-26.}
    \label{fig:gasvismin}
\end{figure*}

\begin{figure*}[htb!]
    \plotone{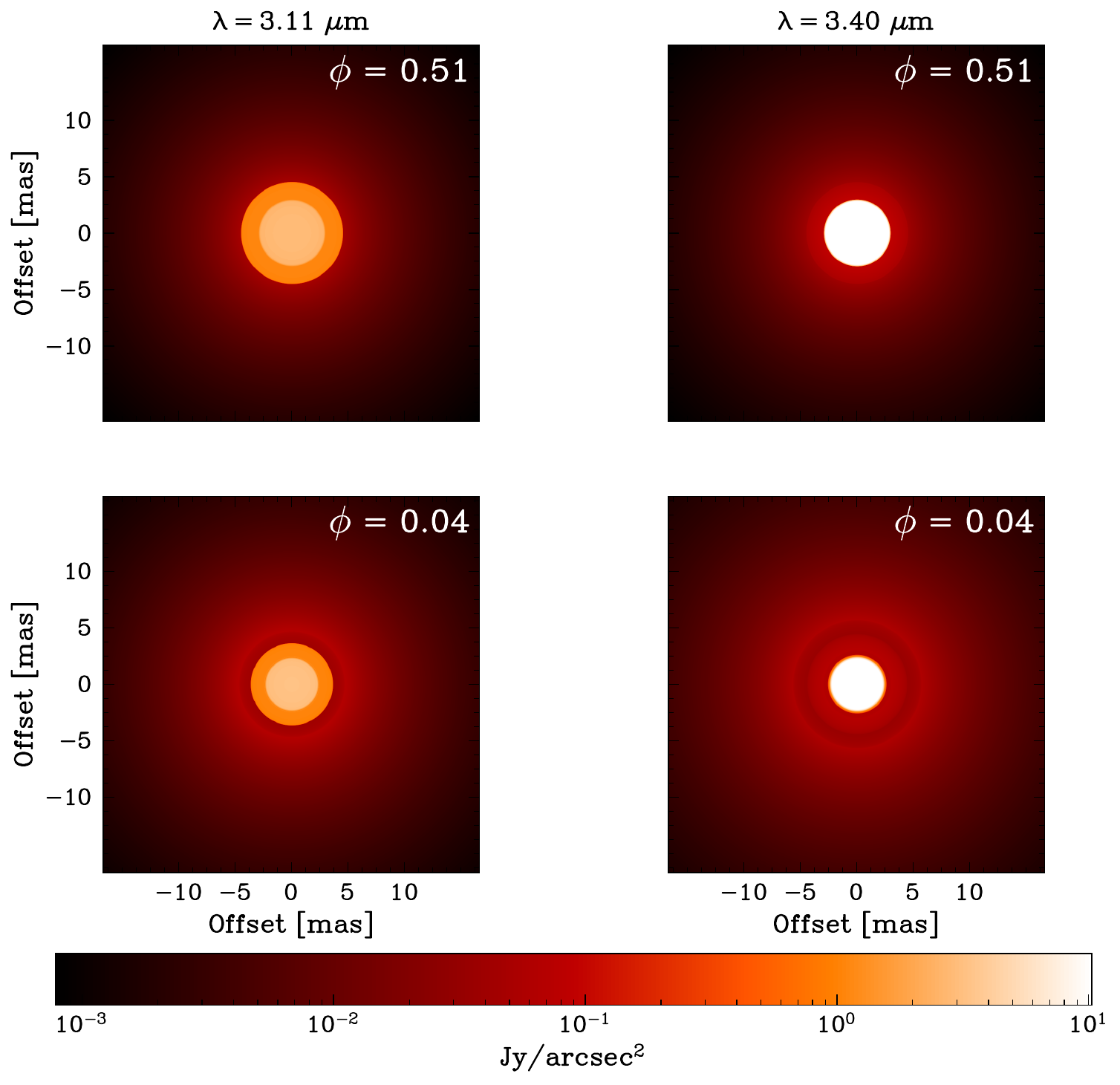}
    \caption{RADMC-3D images of the best fitting model at the minimum of the V-band light curve (top row) and maximum light model (bottom row) at the $3.1~\mu$m feature and the pseudocontinuum at $3.4~\mu$m. The images are convolved to the MATISSE spectral resolution of 34.}
    \label{fig:imagesL}
\end{figure*}
\section{Results}\label{sec:Results}

\subsection{Geometric Modeling}\label{sec:geomres}
The uniform disk (UD) diameter for each night is reported at $3.1~\mu$m, $3.4~\mu$m, and $8.5~\mu$m in Table \ref{tab:UDtable}. For the binned observations with phase 0.04. The UD diameter is $10.07~\pm~0.4~mas$ with reduced $\chi^2=3.3$ at $3.1~\mu$m and $7.8~\pm~0.2~mas$ with reduced $\chi^2=0.8$. For the binned phase of 0.51, the UD diameter is $10.1~\pm~0.2~mas$ with reduced $\chi^2=1.0$ at $3.1~\mu$m and $6.5~\pm~0.1~mas$ with reduced $\chi^2=0.1$. 

\subsection{RADMC-3D}\label{sec:radres}
\begin{figure*}[tb]
    \plotone{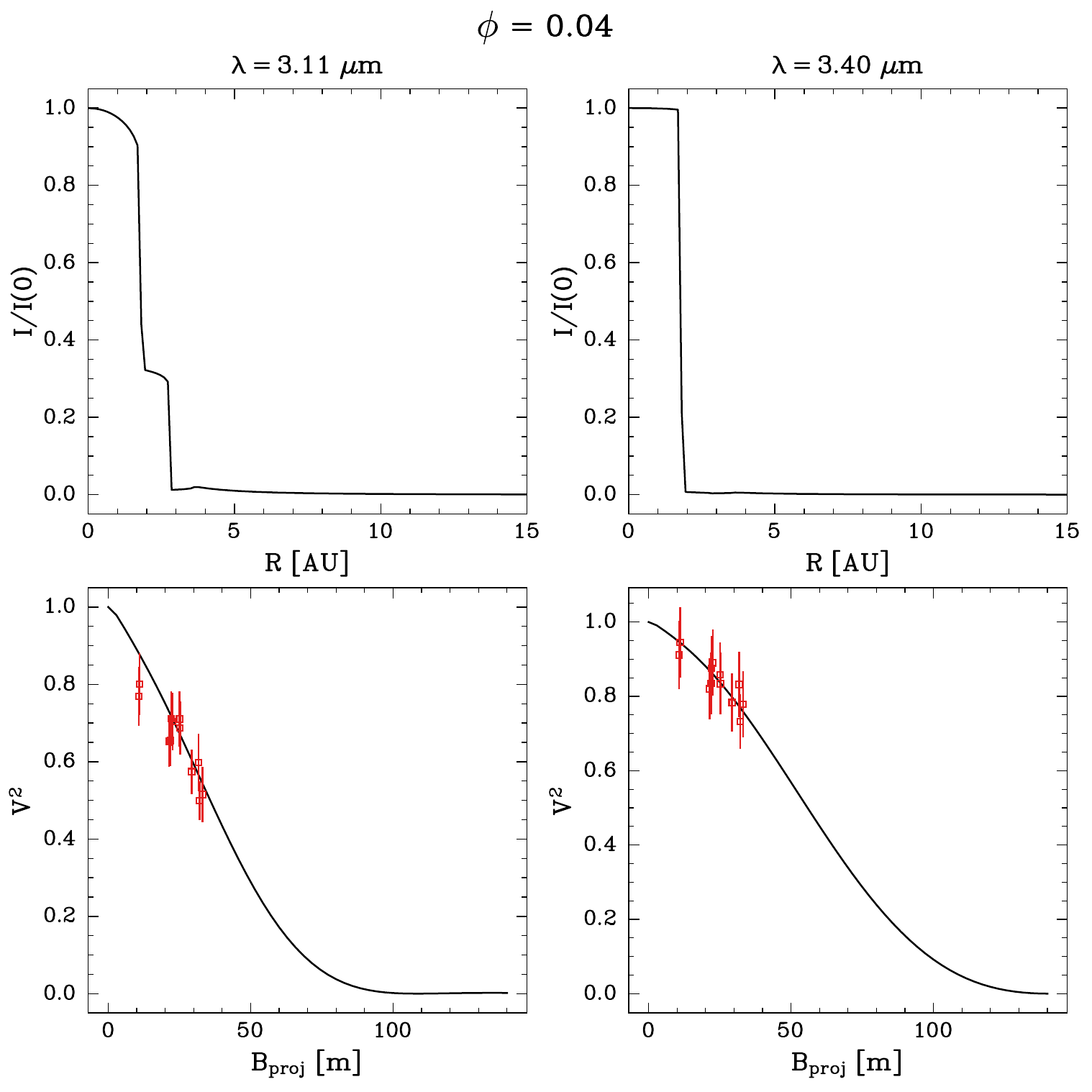}
    \caption{ RADMC-3D radial profiles of the best fitting model at the maximum of the V-band lightcurve for wavelengths $3.1$ and $3.4~\mu$m (top row) and plots of the model visibility vs baseline compared to the observed visibilities (bottom row).}
    \label{fig:radprofmax}
\end{figure*}

\begin{figure*}[tb]
    \plotone{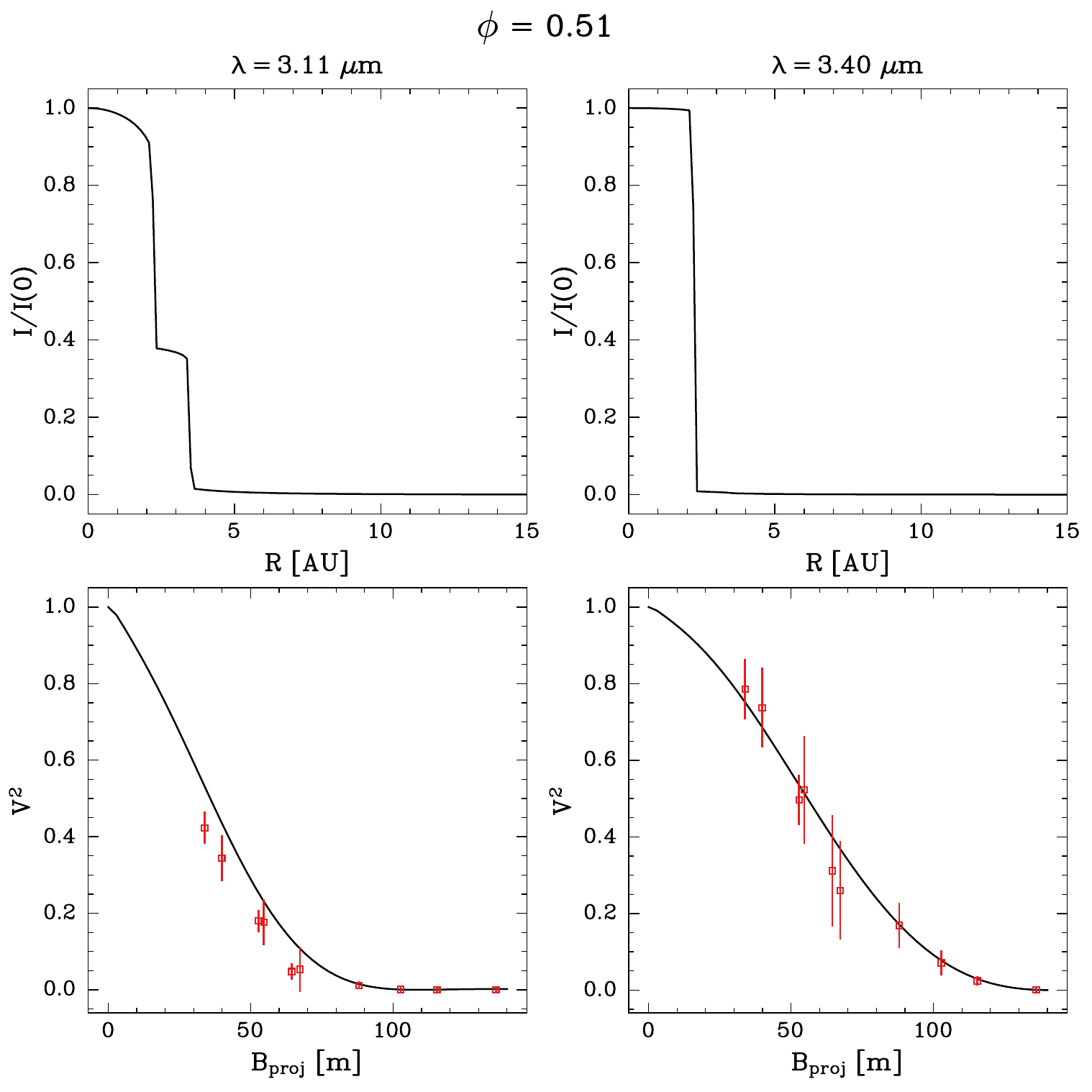}
    \caption{Same as Figure~\ref{fig:radprofmin} but at $\phi~=~0.51$. The two longest baselines from the night of 2022-08-28 are excluded as noted in Section \ref{sec:Observations}.}
    \label{fig:radprofmin}
\end{figure*}

We show the results of the best fitting model from the dust shell only modeling in Table~\ref{tab:dustshellparameters}. Because of the computation time required to fully calculate line radiative transfer in the near-infrared, we only added the molecular shell to the best fitting models from this first step.  We list the best fitting models from the second step of the modeling process in Table~\ref{tab:gashellparameters}, and their L-band visibility profiles are plotted in Figures~\ref{fig:gasvismax} and \ref{fig:gasvismin} for the nights of 2022-03-12 and 2022-08-26. Images at $3.11$ and $3.4~\mu$m are shown in Figure~\ref{fig:imagesL} and the radial intensity profiles at the same wavelengths are shown in Figures~\ref{fig:radprofmax} and \ref{fig:radprofmin}. Plots models compared to the L-band observations on 2022-03-12, 2022-08-24, and 2022-08-28 are shown in Appendix~\ref{sec:appextra}. We do not present a plot comparing the models to the single M-band observation because the contributions from CO in the M-band are not accounted for in our modeling. The L-band spectra of carbon stars are dominated by HCN and C$_2$H$_2$ features  (\citealp{Matsuura2002}), which make it difficult to constrain the distribution of CO accurately.  Uncertainties in the model parameters are not reported since we employed a grid search strategy to optimize the dust shell parameters.

\begin{deluxetable*}{lllllll}[htb]\label{tab:gashellparameters}
\tablehead{\colhead{}&\colhead{Photospheric Radius}&\colhead{Dust Radius}&\colhead{Gas Shell Width}&\colhead{log N(C$_2$H$_2$)}&\colhead{log N(HCN) }&\colhead{L-Band}\\[-0.25cm]
\colhead{Phase}&\colhead{($R_{\odot}$)}&\colhead{($R_*$)}&\colhead{($R_*$)}&\colhead{$cm^{-2}$}&\colhead{$cm^{-2}$}&\colhead{$\chi^2_{red}$}}
\tablecaption{Parameters for dust+gas shell models.}
\startdata
        0.04 & 395 &2.0 & 1.5 & 22.8 & 20.2& 0.43\\
        0.51 & 495 &1.5 & 1.5 &22.6 & 19.9 & 0.48\\
\enddata
\end{deluxetable*} 

For both the minimum and maximum light best fitting models, the overall shape is flat aside from a large decrease in visibility levels around $3.11~\mu$m.  This is consistent with the larger apparent diameter observed in the uniform disk fits in Figure~\ref{fig:pmoirednobin}.  There is also a smaller decrease in the visibility levels  at wavelengths longer than $3.5~\mu$m is visible in the fits at the light curve minimum, but not at the light curve maximum. The visibility levels are high at projected baselines of $\sim30~m$, but fall off quickly with projected baseline. At baselines above $100~m$, the model is mostly resolved out. The reduced $\chi^2$ is fair at both epochs, but has a higher value of 3.4 at the visual light curve minimum ($\phi~=~0.51$). This is due in part to the fact that more of the observations were resolved out with small errorbars at $\phi~=~0.51$, and also caused by the L-band observations on 2022-08-28 (Figure~\ref{fig:Ldata}). The highest two projected baselines of $95.9~m$ and $104.1 ~m$ show a deviation from the overall flat visibility profile at wavelengths longer than $3.5~\mu$m, but were not included in the calculation of the $\chi^2$ as they were affacted by the callibration issue discussed in Section \ref{sec:Observations}.
\section{Discussion}\label{sec:Discussion}
\subsection{Geometric Modeling}\label{sec:geomdisc}
At both epochs, the uniform disk $\chi^2$ is lower at the pseudocontinuum of $3.4~\mu$m than at the $3.11~\mu$m molecular feature, indicating the pseudocontinuum is better modeled by the uniform disk assumption than the molecular shell. The angular size estimates in the L-band at $3.1~\mu$m and the N-band at $8.5~\mu$m correspond to the extent of the molecular shell composed of C$_2$H$_2$ and HCN. The $3.4~\mu$m diameter estimate falls in the pseudo-continuum where the molecular contributions are less significant, but may not provide an uncontaminated estimate of the photospheric size. The M-band diameter estimate is also affected by a CO spectral features, including at $4.78~\mu$m, making the L-band pseudocontinuum the best indicator of the size of the star. In the M-band data, there is a drop the visibility at $4.78~\mu$m, with a corresponding increase in the uniform disk diameter. However, it is not clear if the observed feature is due to CO, because the uncertainty of the visibility at $4.78~\mu$m is large compared to the uncertainties of the surrounding points. 
\subsection{Visibility Modeling}
The $\phi~=~0.04$ model fits the visibility data well, with a reduced $\chi^2$ of 0.43. However, we could not reproduce exactly the shape of the $3.11~\mu$m molecular feature. The depth of the feature is overall reproduced well by the models, but the model feature's center is consistently at a shorter wavelength than observed in the data. Additionally, the modeled $3.11~\mu$m feature is consistently wider than the width of the observed feature. At $3.0~\mu$m, the observed visibility is flat, whereas the modeled visibility begins to decrease around $2.9~\mu$m. One possible reason for these discrepancies is that we used a flat gas temperature distribution of $1200~K$. In nature, the gas temperature is likely to be higher closer to the star, and lower further away from it. RADMC-3D determines the dust temperature through Monte-Carlo random walk methods, but it requires the gas temperature to be provided by the user. Fitting a more complex profile was beyond the scope of this article, and requires a high computation time to simulate the gas shell contributions. Changing the uniform temperature of the gas primarily affects the width of the feature, and does not affect the position of the center of the $3.11~\mu$m feature. It is also possible that the gas around V Oph is not in local thermal equilibrium. Modeling non-LTE effects is also beyond the scope of this article. 
The $\phi~=0.04$ model has a dust radius of $2.0~R_*$ where $R_*$ corresponds to a stellar radius of $395~R_{\odot}$. The outer-radius of the gas shell only extends to $1.5~R_*$, indicating that in this model, the most pronounced gas shell does not overlap with the dust shell. As more parameter space is explored, the possibility remains that a better model incorporating the co-existence of dust and gas could be found. The minimum light curve model at $\phi=0.51$ has a reduced $\chi^2$ of 0.51. The model displays the same general trends as the maximum V-band light curve model. The $3.11~\mu$m feature's width and central wavelength do not match the observations in this observational epoch as well. An additional discrepancy between model and data comes from the observed decrease in the visibility levels beginning around $3.7~\mu$m, which the model does not reproduce. increasing the number density of HCN increases the depth of this feature in the model, but it also causes a drop in visibility around $3.5~\mu$m, which is not observed in the data. 

In the $\phi=0.51$ model, the dust radius is localized at $1.5~R_*$. However, the stellar radius $R_*$ is $495~R_{\odot}$. This means the dust radius of $742.5 R_{\odot}$ is similar to the dust radius of $790~R_{\odot}$ derived for the phase of $\phi=0.04$. The edge of the gas shell lies at the dust radius of $1.5~R_*$. This indicates the dust and most prominent gas shell can be co-located at least for parts of the pulsational period of the star. 

\subsection{Dust Temperature}
RADMC-3D calculated the dust temperatures of each model using a random walk algorithm. The amC dust temperature of the maximum V-band light model at the dust radius of 2.0~$R_*$ is 1540~K, and the dust temperature of the minimum light model at the dust radius of 1.5~$R_*$ is 1740~K. In a review of AGB mass loss, \cite{Hofner2018} cite a typical amC dust condensation temperature of 1500~K.  Other modeling efforts such as \citet{Sacuto2011} and \citet{Rau2015} used dust condensation temperatures of 1200 K and 1000-1300 K, respectively.  \citet{OH7} found an inner radius dust temperature of 1580 K. RADMC-3D does not account for dust condensation chemistry, and therefore  does not require the temperature at the dust condensation radius to be consistent across models, nor does it need to align with any specific canonical condensation temperature.

\subsection{Radial Profiles}
The drop in visibility level at the $3.11~\mu$m feature is larger in the minimum V-band lightcurve epoch of $\phi=0.51$ than at the maximum V-band epoch ($\phi=0.04$). The model radial profiles in Figures~\ref{fig:radprofmax} and \ref{fig:radprofmin} combined with the simulated images in Figure~\ref{fig:imagesL} provide an intuitive interpretation of this phenomenon. At $3.11~\mu$m, the $\phi=0.04$ radial profile shows a smooth decrease in intensity out to the stellar radius around 2 AU. Over the next AU, the intensity remains flat due to gas emission until the edge of the gas shell is reached, and it drops sharply again. In the $\phi=0.51$ profile, the same pattern is observed, but the emission due to the gas shell is higher relative to the emission at the center of the star. Because of the higher level of extended emission around the central source, the corresponding visibility levels in the $3.11~\mu$m feature are lower at $\phi=0.51$. At both epochs, the $3.40~\mu$m shows lower levels of gas emission, so there is a smaller drop in the level of the visibility.

\subsection{Comparison to Rau et al. 2019}
\cite{RAU2019} fit MIDI observations using the dynamic atmosphere models described in \cite{DMA}, and reported stellar and dust radii of the best fitting model, enabling a comparison to the models in this work. \citet{RAU2019} adopted the same binning used by \citet{OH7}, with data grouped into three epochs of phase $\phi=0.18,~0.49,~0.65$. In Table \ref{tab:compare}, the phase of 0.49 is directly compared to our minimum light curve data with $\phi=0.51$. We also compare with caution the $\phi=0.16$ data to our maximum light curve epoch of $\phi=0.04$. 

\begin{deluxetable*}{llllll}\label{tab:compare}
\tablehead{\colhead{}&\colhead{Rau et al. 2019 $R_*$}&\colhead{Rau et al. 2019 $R_{dust}$}&\colhead{}&\colhead{$R_*$}&\colhead{$R_{dust}$}\\[-0.2cm]
\colhead{MIDI $\phi$}&\colhead{($R_{\odot}$)}&\colhead{($R_{\odot}$)}&\colhead{MATISSE $\phi$}&\colhead{($R_{\odot}$)}&\colhead{($R_{\odot}$)}}
\tablecaption{Stellar parameters from \cite{RAU2019} compared to our modeling.}
\startdata
        0.18 & 479 &780& 0.04 & 395 & 790\\
        0.49 & 494 &853& 0.51 & 495 & 742.5\\
\enddata
\label{tab:compare}
\end{deluxetable*}
The stellar radius from \citet{RAU2019} is highest at the minimum of the V-band light curve,  following the same trend as our modeling. However, the phase of 0.18 has a stellar radius of $479~R_{\odot}$, which is significantly larger than our value of $395~R_{\odot}$. Because of observational constraints, no long baseline MATISSE data was obtained near the light curve maximum -  if that data was obtained, a larger stellar radius may be favored. One piece of evidence supporting this idea is the good fit of the best fitting minimum light curve dust shell only model to the maximum light data. The reduced $\chi^2$ of this model is 1.06, compared to the best fitting model $\chi^2$ of 0.40. This indicates that a wide range of models, including those with larger stellar radii may fit the data at $\phi=0.04$. Further observations of V Oph with appropriate baseline coverage in the L-band will help to break this degeneracy. The stellar radii from both papers agree to within $1~R_{\odot}$ at the visual light curve minimum. At the visual light curve maximum, the dust radii are similar, while they differ by about $100~R_{\odot}$ at the visual light curve minimum. We note the dust radius derived in \cite{RAU2019} has a value of $1.79~R_*$. Our simulated grid of models had a spacing $0.5~R_*$ in dust radius, so the value of \cite{RAU2019} is consistent within the resolution of our simulations. As such, our L-band observations and radiative transfer modeling are consistent with the N-band hydrodynamic modeling of \cite{RAU2019}.

The trend of a larger stellar radius at the light curve minimum than near the maximum has also been observed in oxygen-rich AGB stars. For example, \citet{RPEG} reported snapshot observations of the O-rich Mira R Peg, and found the same trend of larger radius at minimum light.

\subsection{Interferometric Variability \& Imaging }
\begin{figure}[htb!]
    \plotone{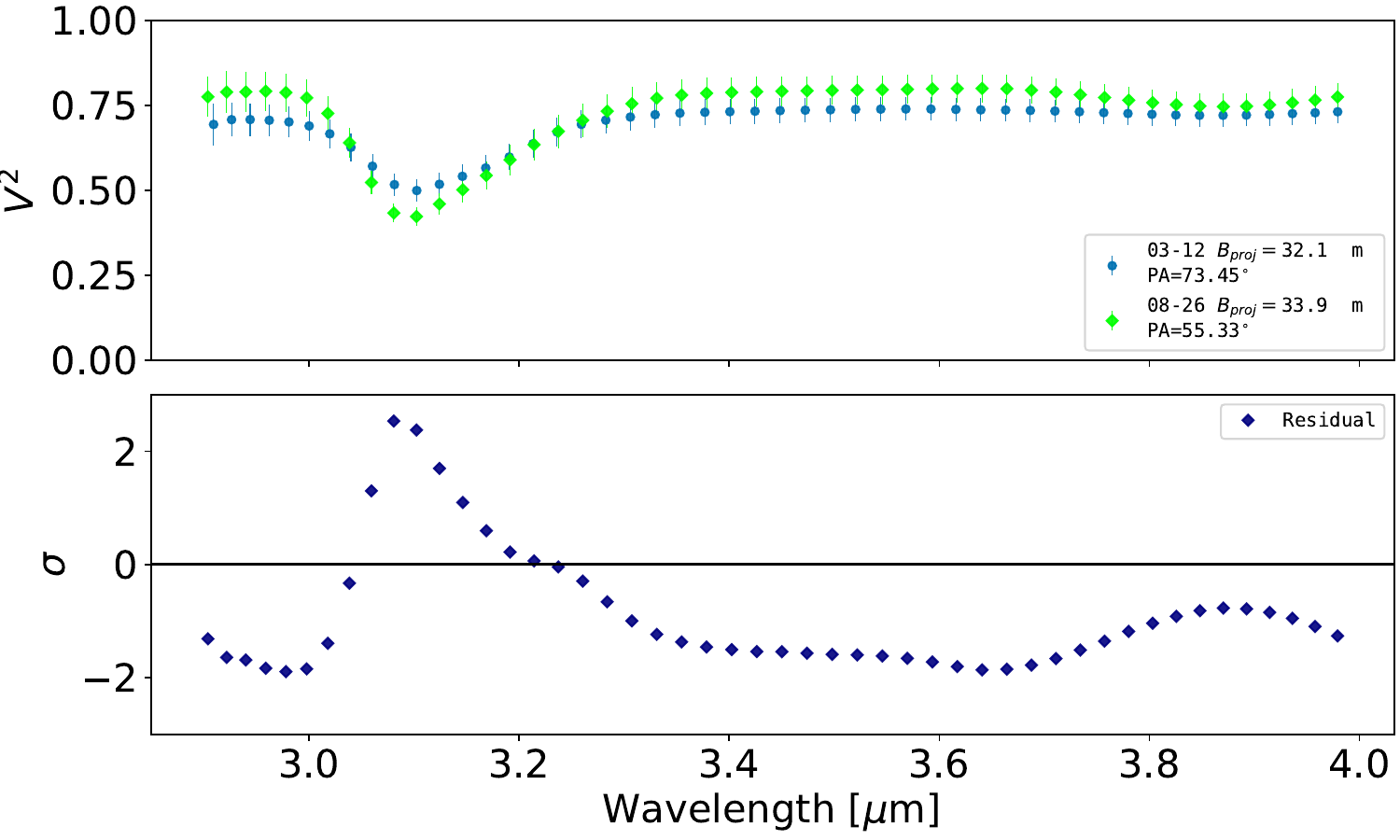}
    \caption{Top - comparison of similar baselines and position angles near maximum and minimum light. Variability of up to 20\% is observed in the 3.11 $\mu$m feature. Bottom -difference between visibility profiles scaled by the errorbars of the 2022-03-12 observations.}
    \label{fig:VarComp}
\end{figure}

\begin{figure*}[htb!]
    \plotone{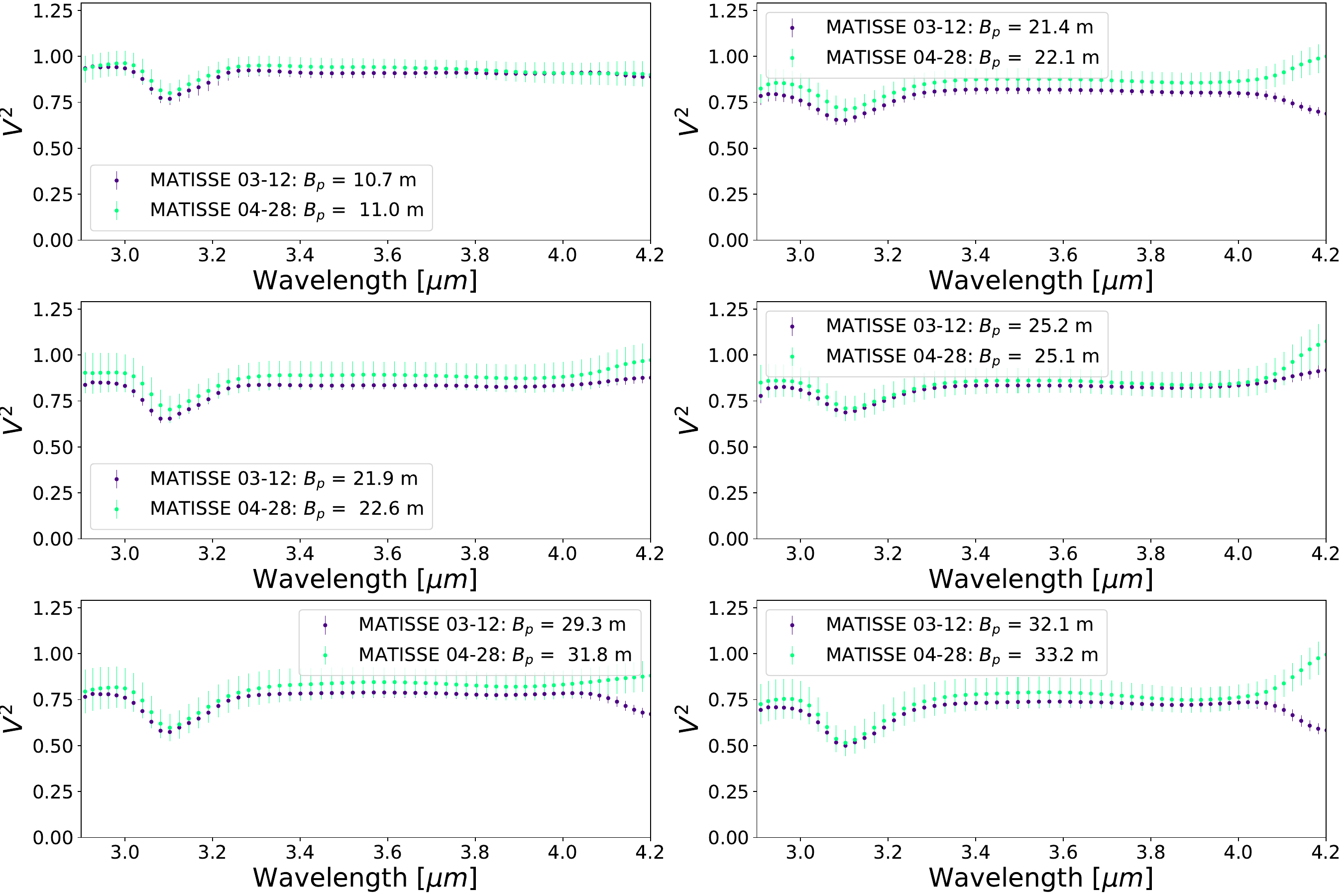}
    \caption{Comparison of observations separated by 47 days, or approximately 20\% of the V Oph period. }
    \label{fig:VarComp2}
\end{figure*}

Interferometric variability is defined as a change in the interferometric observables of a target over time. Given the rapidly changing extended atmosphere and pulsation driven dust formation of AGB stars, it is often observed in stars such as V Oph. The MIDI observations from \cite{OH7} show evidence of interferometric variability in the N-band. To evaluate the the level of interferometric variability in the L-band, we compared the MATISSE observations at minimum and maximum light. Near maximum light, two observations (03-12-2022) were made in the small VLTI configuration. Near minimum light, four observations (08-26-2022) were made in medium and large configurations. Because the observations were made in different configurations, direct comparisons can only be made between baselines of approximately 30 m. Figure \ref{fig:VarComp} shows a comparison of the 30 m baselines at maximum and minimum light for each wavelength channel. The difference in visibility is most pronounced at the $3.11~\mu$m feature. The observation from minimum light has a deeper drop in visibility at $3.11~\mu$m, with a maximum percent difference in visibility level of 20, and a median percent difference difference of 7. The median difference scaled by the errors from the night of 03-12-2022 is 2.0. This indicates that for some wavelength channels, there is measurable interferometric variability between minimum and maximum light.  To evaluate the interferometric variability on shorter timescales, we compare the observations obtained on 04-28-2022 to those made on 03-12-2022. These observations are separated by 47 days, approximately two tenths of the V-band light curve period of the V Oph. Both of these observations were made in the small configuration and have similar baselines and position angles, so it is possible to directly compare the visibility profiles observed on each date. As shown in Figure \ref{fig:VarComp2}, the median difference difference between the visibility profiles on the two dates scaled by the errors from 2022-03-28 is 1.3. As we discussed in Section~\ref{subsec:visdata}, this suggests the observations are consistent with each other with no sign of interferometric variability.

The level of variability is relevant to planned interferometric imaging efforts of V Oph. Because variability is observed over a timescale of half the period (roughly 150 days), but is not significant on timescales of roughly 20\% of the period (50 days), imaging observations of V Oph conducted within two months of each other may be combined without being affected by the interferometric variability.
\section{Conclusions}\label{sec:conclusions}
We have described new L-, M-, and N-band observations of V Oph with the MATISSE instrument and modeled them using geometric fitting and radiative transfer modeling. The geometric fitting at the $3.11~\mu$m C$_2$H$_2$ and HCN feature shows evidence of an extended molecular layer around the star. With RADMC-3D modeling, we determined a stellar radius 395 and $495~R_{\odot}$ and a dust radius of 790 and $742.5~R_{\odot}$ at the two phases of $\phi=0.04$ and $\phi=0.51$.   We modeled the gas shell around the star by including C$_2$H$_2$ and HCN in the simulations, and found the $\phi=0.04$ data is described by a gas shell that extends to $585~R_{\odot}$, and the $\phi=0.51$ model gas shell extends to $743~R_{\odot}$. The model visibilities agree well with the observations, with deviations around the molecular features at $3.11~\mu$m and $3.8~\mu$m.

Further observations of V Oph are necessary because a wide range of model parameters can explain the data at $\phi=0.04$, for which all projected baselines are less than $40~m$. Longer baselines will break this degeneracy, and can be obtained as part of an imaging program. It will be possible to image V Oph because the observed interferometric variability is negligible on timescales of less than 40 days. Imaging observations will break the aforementioned degeneracy and can be used to probe the inner regions of the extended atmosphere for asymmetries. Imaging V Oph in the K-band with GRAVITY \citep{GRAV} as well as in the L-band with MATISSE will allow us to further constrain the shape of the molecular shell by probing the distribution of CO molecules.

The modeling process can be improved by including a way to physically determine the gas temperature, or to efficiently sample a range of gas temperatures. One way to speed up the computation time would be to replace the line radiative transfer of RADMC-3D with pre-calculated opacities such as those provided by the Opacity Project \citep{opacity}.

%




\begin{acknowledgments}
This material is based upon work supported by NASA under award numbers 80GSFC21M0002 and 80NSSC24M0022. GR acknowledge the support received for performing this work while serving at the National Science Foundation. We acknowledge with thanks the variable star observations from the AAVSO International Database contributed by observers worldwide and used in this research. This
work has made use of data from the European Space Agency
(ESA) mission Gaia (https://www.cosmos.esa.int/gaia), pro-
cessed by the Gaia Data Processing and Analysis Consortium
(DPAC, https://www.cosmos.esa.int/web/gaia/dpac/
consortium). Funding for the DPAC has been provided by
national institutions, in particular the institutions participating in the Gaia Multilateral Agreement. NSO/Kitt Peak FTS data used here were produced by NSF/NOAO. We thank the anonymous referee for their constructive comments. 
\end{acknowledgments}
\facility{VLTI, AAVSO}
\software{PMOIRED, RADMC3D}

\appendix
\section{Additional Model Plots}\label{sec:appextra}
\begin{figure*}[ht!]
    \plotone{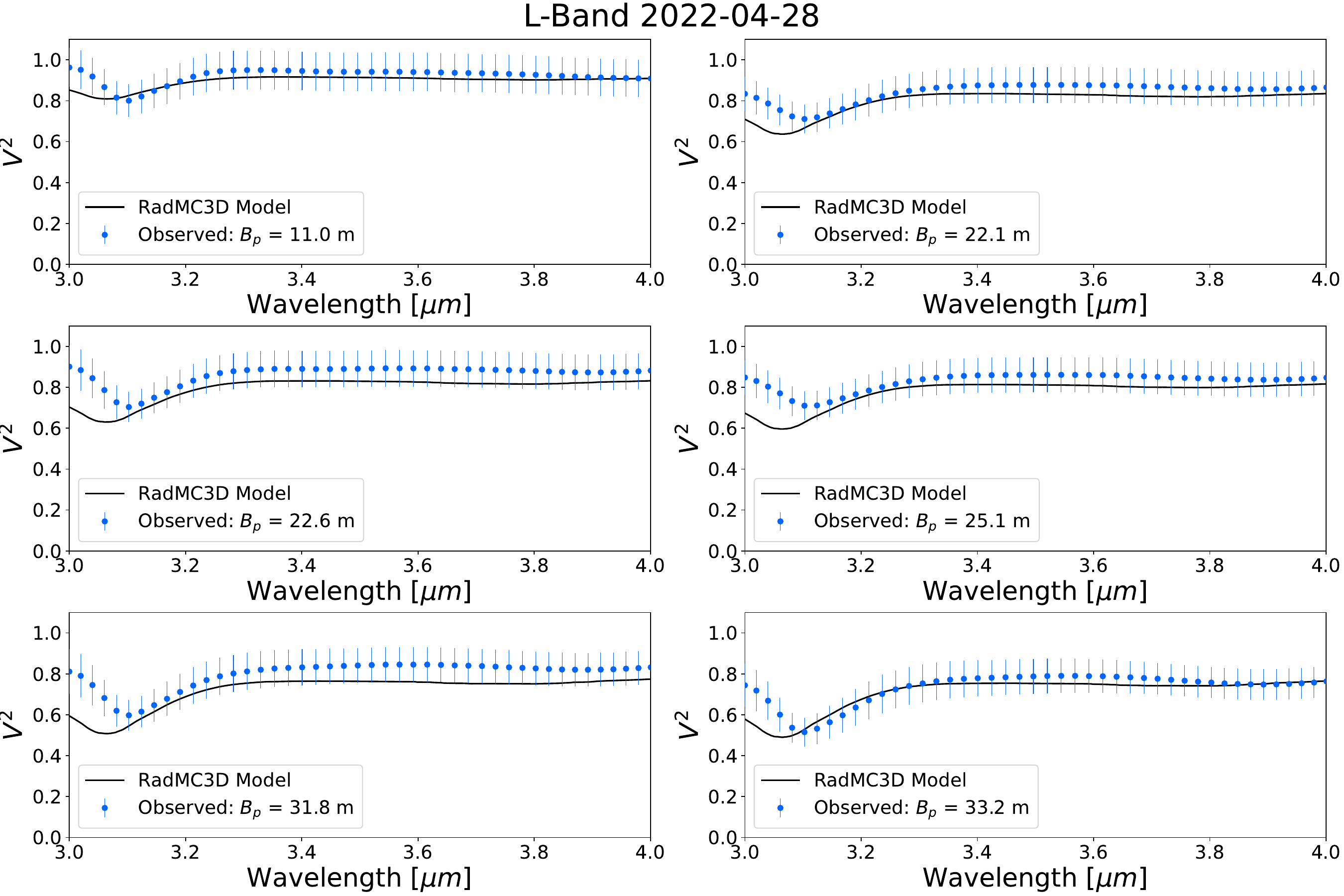}
    \caption{Synthetic RADMC3D-calculated (black lines) and observed (red dots) L-band visibility profiles for the maximum light epoch on the night of of 2022-04-28.}
    \label{fig:gasvismax0428}
\end{figure*}

\begin{figure*}[ht!]
    \plotone{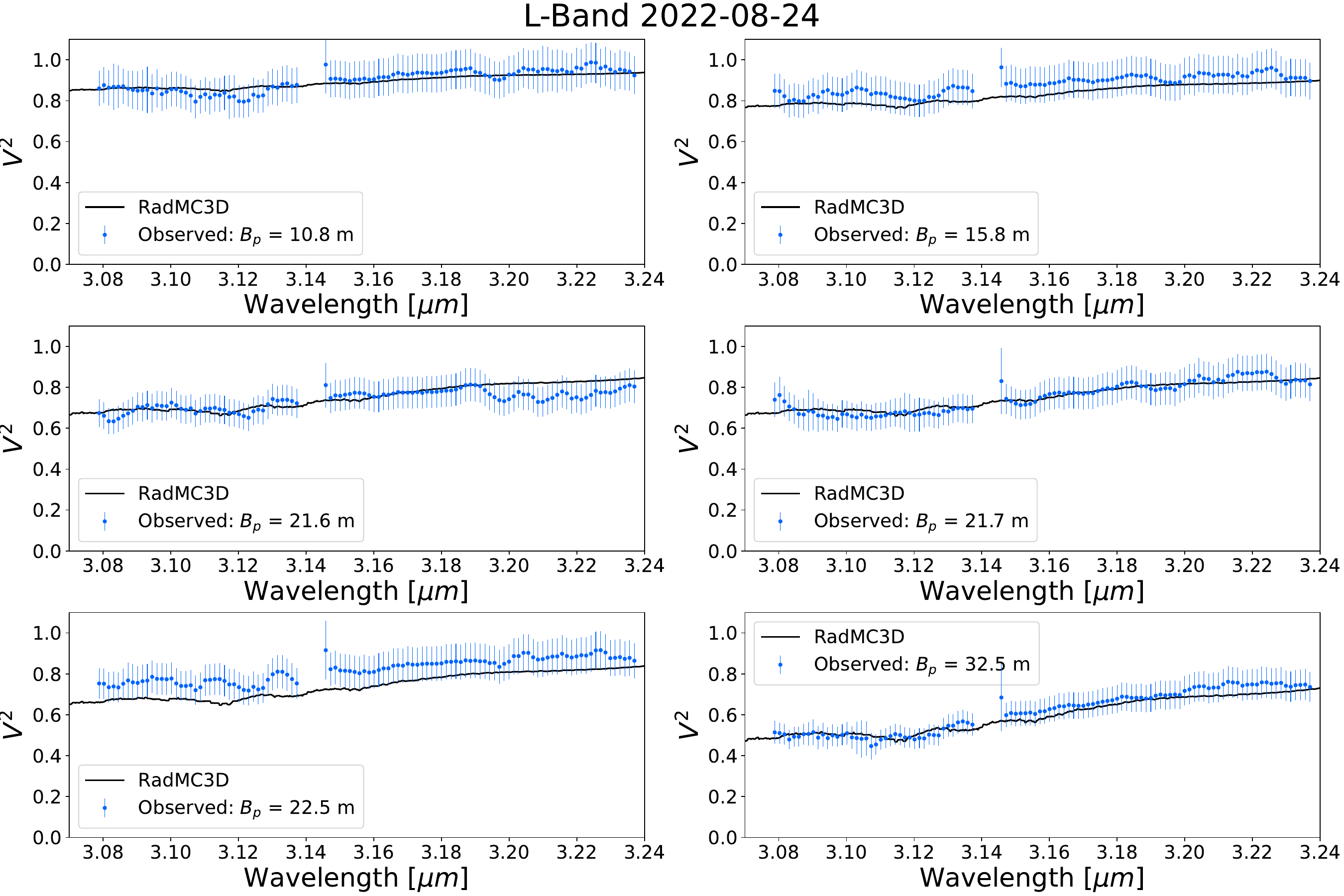}
    \caption{Synthetic RADMC3D-calculated (black lines) and medium resolution L-band visibility profiles on the night of 2022-08-24 ($\phi=0.51$).}
    \label{fig:gasvismin0824}
\end{figure*}

\begin{figure*}[ht!]
    \plotone{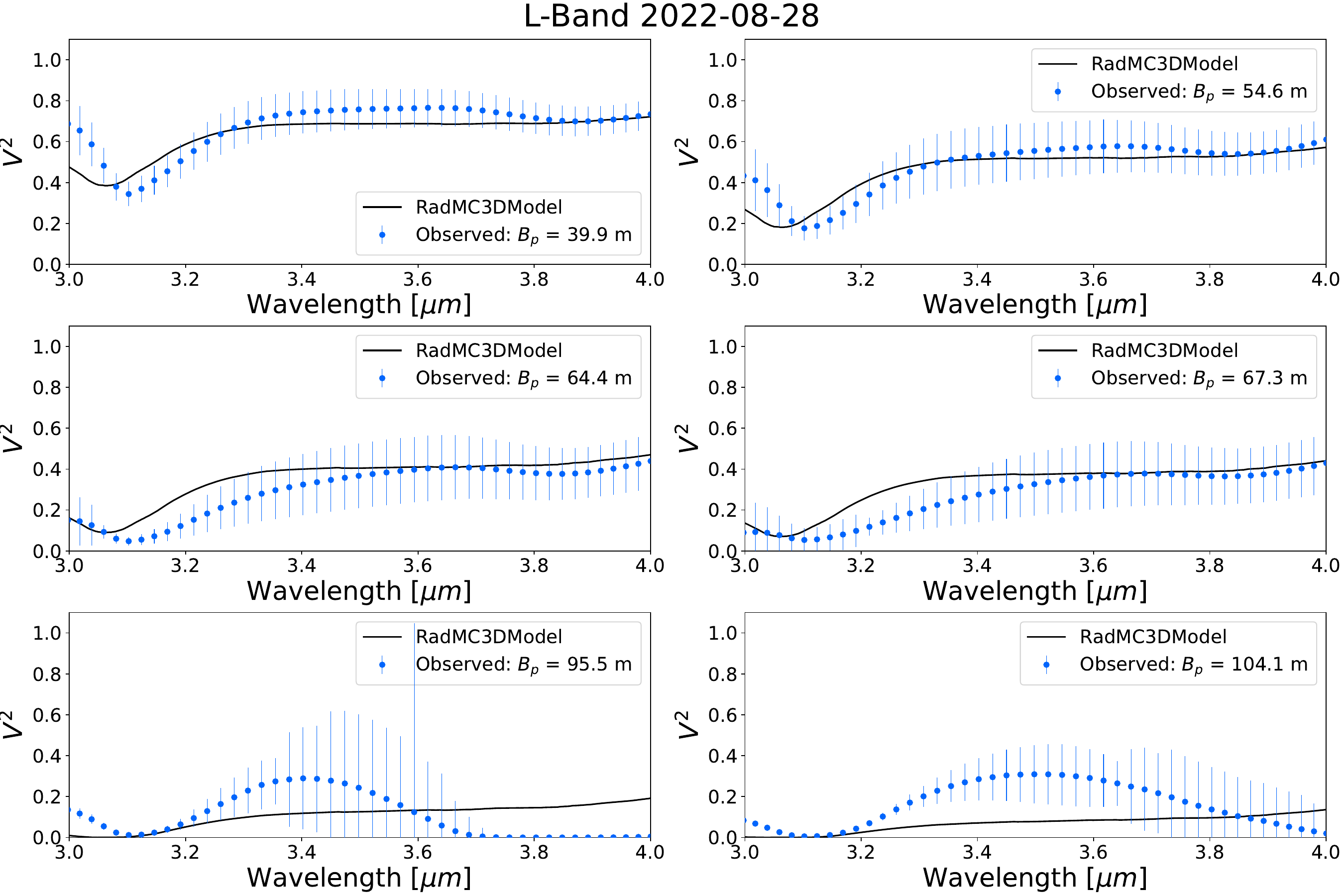}
    \caption{Same as Figure \ref{fig:gasvismax0428} but for the the night of 2022-08-28.}
    \label{fig:gasvismin0828}
\end{figure*}

Figures~\ref{fig:gasvismax0428},~\ref{fig:gasvismin0824}, and~\ref{fig:gasvismin0828} show the model comparison to the three nights of L-band observations not shown in Section~\ref{sec:radres}: 2022-04-28, 2022-08-24, and 2022-08-28. The observations from the night of 2022-08-24 are the sole night of observations obtained with medium spectral resolution (R=506).

\begin{figure*}[ht!]
    \plotone{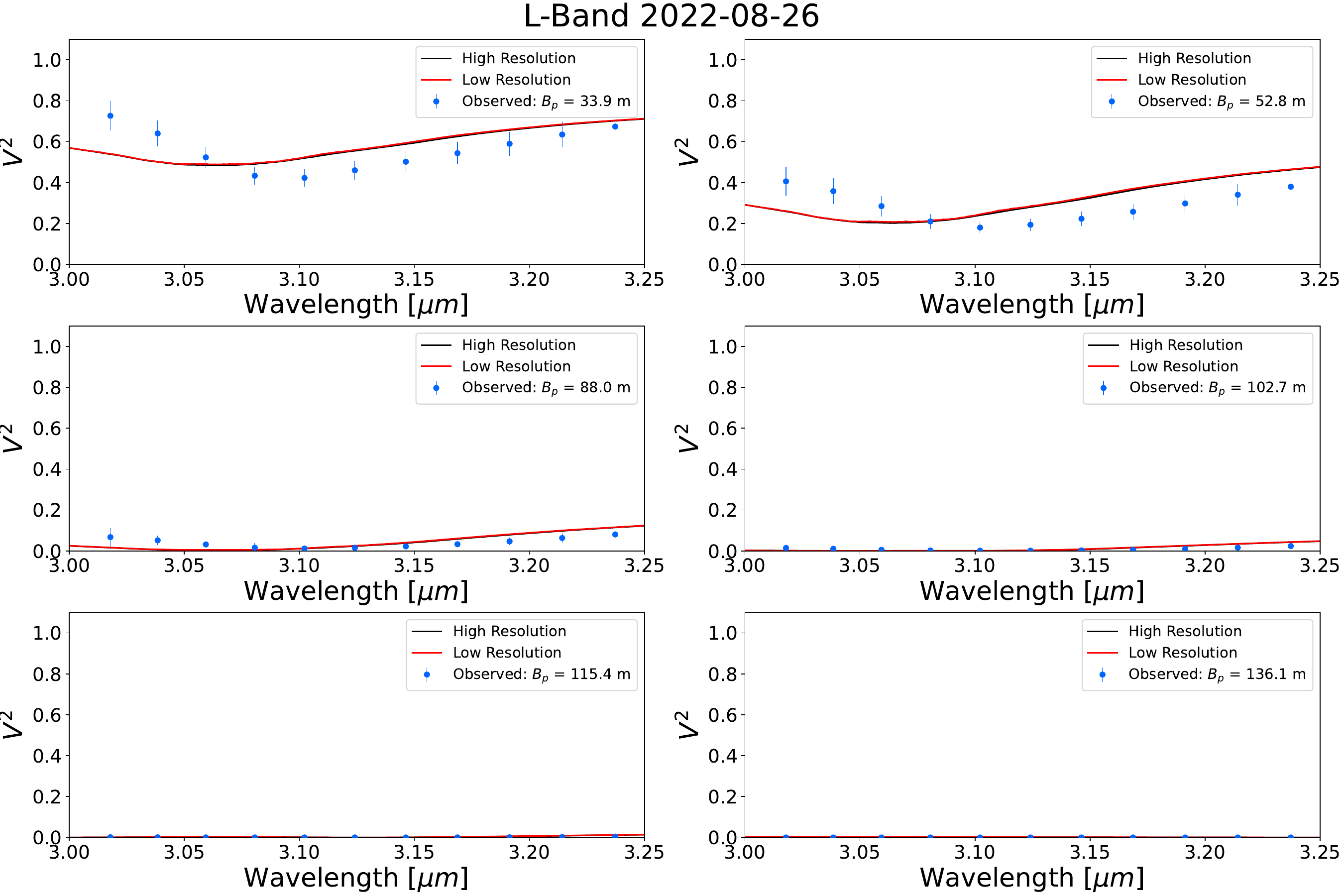}
    \caption{Comparison of synthetic visibilities for low spectral resolution (red) and high spectral resolution simulations (black) on the night of 2022-08-26.}
    \label{fig:compareLRHR}
\end{figure*}

We also include a comparison of the low spectral resolution and high resolution synthetic visibilities for the minimum light model on the night of 08-26-2022 in Figure \ref{fig:compareLRHR}. The difference between the low- and high-resolution models is small. The reduced $\chi^2$ of the low spectral resolution maximum V-band light model is 0.43, the same as the reduced $\chi^2$  of high spectral resolution model. The reduced $\chi^2$ of the low spectral resolution minimum light model is 0.5, compared to 0.48 for the minimum light high spectral resolution model.

\clearpage

\bibliography{main}{}

\bibliographystyle{aasjournal}



\end{document}